\documentclass{aa}  
\usepackage{graphicx}
\usepackage{txfonts}
\newcommand{\CI}{[C \textsc{i}]}
\newcommand{\OI}{[O \textsc{i}]}
\newcommand{\thetaonec}{$\theta^1$ Ori C}
\newcommand{\nodata}{...}

\begin{document}

   \title{{A tell-tale tracer for externally irradiated protoplanetary disks: Comparing the [C\textsc{i}]~8727~\AA\ line and ALMA observations in proplyds} \thanks{Based on observations collected at the European Southern Observatory under ESO programme 0104.C-0963(A), 106.218X.001, and 110.259E.001.}}
    \titlerunning{A tell-tale tracer for externally irradiated protoplanetary disks}
   \author{M.-L. Aru\inst{\ref{instESO}}
          \and
          K. Mauc\'o\inst{\ref{instESO}}
          \and 
          C. F. Manara\inst{\ref{instESO}}
          \and
          T. J. Haworth\inst{\ref{instQMUL}}
          \and
          N. Ballering\inst{\ref{instVirg}}
          \and
          R. Boyden\inst{\ref{instVirg},\ref{instSpaceSci}}
        \and
          J. Campbell-White\inst{\ref{instESO}}
          \and 
          S. Facchini\inst{\ref{uniMI}}
          \and 
        G. P. Rosotti\inst{\ref{uniMI}}
          \and 
        A. Winter\inst{\ref{instOCA},\ref{instIPAG}}
        \and
         A. Miotello\inst{\ref{instESO}}
          \and
          A. F. McLeod\inst{\ref{Durham},\ref{Durham2}}
          \and
          M. Robberto\inst{\ref{STScI},\ref{JHU}}
          \and 
          M. G. Petr-Gotzens\inst{\ref{instESO}}
        \and
          G. Ballabio\inst{\ref{instImperial}}
          \and \\
          S. Vicente\inst{\ref{instPO}}
          \and 
          M. Ansdell\inst{\ref{NASA}} 
          \and
     L. I. Cleeves\inst{\ref{instVirg}}
          }

   \institute{European Southern Observatory, Karl-Schwarzschild-Strasse 2, 85748 Garching bei München, Germany\label{instESO}\\
              \email{mariliis.aru@eso.org}
          \and
          Astronomy Unit, School of Physics and Astronomy, Queen Mary University of London, London E1 4NS, UK\label{instQMUL}
          \and
          Department of Astronomy, University of Virginia, Charlottesville, VA 22904, USA\label{instVirg}
          \and
           Space Science Institute, Boulder, CO 80301, USA\label{instSpaceSci}
      \and
          Dipartimento di Fisica, Università degli Studi di Milano, Via Celoria 16, 20133 Milano, Italy\label{uniMI}
          \and
          Centre for Extragalactic Astronomy, Department of Physics, Durham University, South Road, Durham DH1 3LE, UK\label{Durham}
          \and
          Department of Physics, Institute for Computational Cosmology, University of Durham, South Road, Durham DH1 3LE, UK\label{Durham2}
          \and
          Space Telescope Science Institute, 3700 San Martin Dr, Baltimore, MD 21218, USA\label{STScI}
          \and
          Johns Hopkins University, 3400 N. Charles Street, Baltimore, MD 21218, USA\label{JHU}
          \and
          Instituto de Astrofísica e Ciências do Espaco, Universidade de Lisboa, OAL, Tapada da Ajuda, P-1349-018 Lisboa, Portugal\label{instPO}
          \and
          Universit\'e Cote d'Azur, Observatoire de la Cote d'Azur, CNRS, Laboratoire Lagrange, F-06300 Nice, France\label{instOCA}
          \and
          Universi\'te Grenoble Alpes, CNRS, IPAG, F-38000 Grenoble, France\label{instIPAG}
          \and
          NASA Headquarters, 300 E Street SW, Washington, DC 20546, USA\label{NASA}
              \and
          Imperial Astrophysics, Imperial College London, Blackett Laboratory, Prince Consort Road, London SW7 2AZ, UK \label{instImperial}
         }

   \date{Received 31 July 2024; accepted 25 October 2024}

 
  \abstract
   {The evolution of protoplanetary disks in regions with massive OB stars is influenced by externally driven winds that deplete the outer parts of disks. These winds have previously been studied via forbidden oxygen emission lines, which also arise in isolated disks in low-mass star-forming regions (SFRs) with weak external UV fields in photoevaporative or magnetic (internal) disk winds. It is crucial to determine how to disentangle external winds from internal ones. Here, we report a proxy for unambiguously identifying externally driven winds with a forbidden line of neutral atomic carbon, \CI\,8727~\AA.

    We compare for the first time the spatial location of the emission in the \OI\,5577~\AA, \OI\,6300~\AA, and \CI\,8727\AA{} lines traced by VLT/MUSE-NFM, with the ALMA Band 7 continuum disk emission in a sample of 12 proplyds in the Orion Nebula Cluster (ONC).
    We confirm that the \OI\,5577~\AA{} emission is co-spatial with the disk emission, whereas the \OI\,6300~\AA{} is emitted both on the disk surface and on the ionization front of the proplyds. We show for the first time that the \CI\,8727~\AA{} line is also co-spatial with the disk surface in proplyds, as seen in the MUSE and ALMA data comparison. The peak emission is compatible with the stellar location in all cases, apart from one target with high relative inclination with respect to the ionizing radiation, where the peak emission is located at the disk edge in the direction of the ionizing radiation. 
    
    To verify whether the \CI\,8727~\AA{} line is detected in regions where external photoevaporation is not expected, we examine VLT/X-Shooter spectra for young stars in low-mass SFRs. Although the \OI\,5577~\AA{} and 6300~\AA{} lines are well detected in all these targets, there is a total of $\ll10\%$ detection rate in the case of the \CI\,8727~\AA{} line. This number increases substantially to a $\sim40\%$ detection rate in $\sigma$-Orionis, a region with a higher UV radiation than in low-mass SFRs, but lower than in ONC. 
    
    The spatial location of the \CI\,8727~\AA{} line emission and the lack of its detection in isolated disks in low-mass SFRs strongly suggest that this line is a tell-tale tracer of externally driven photoevaporative winds, which agrees with recent excitation models.}
   \maketitle
%

\section{Introduction}

The evolution of protoplanetary disks is affected by their surrounding environment. In massive clusters, the UV radiation from OB stars heats the gas in nearby protoplanetary disks, and gives rise to externally driven photoevaporative wind. This wind depletes the disk outside-in, causing rapid mass loss and a shorter disk lifetime (see \citealt{WinterHaworth2022} for a review).

Irradiated disks, typically called proplyds, have been studied best in the Orion Nebula Cluster (ONC), exhibiting a teardrop shape with an ionization front as imaged with the Hubble Space Telescope \citep[HST, e.g.,][]{OdellWen1993,bally1998,Ricci2008THENEBULA}, and more recently studied with the Multi-Unit Spectroscopic Explorer (MUSE) instrument in Narrow Field Mode (NFM) at the ESO Very Large Telescope \citep[VLT,][]{kirwan2023,haworth2023,aru2024} and with JWST  \citep{berne23,berne24,habart2024} . 

While external photoevaporation of disks is unique to clusters with massive stars such as the ONC, internal photoevaporative winds due to the high-energy radiation from the central star \citep[e.g.,][]{ercolano2017} may be present in both high-UV environments, and low-mass star forming regions (SFRs) with weak external UV fields. Internal disk winds may also be magnetohydrodynamically driven (see \citealt{pascucci2023} for a review).
The forbidden emission lines, such as \OI~5577~\AA{} and \OI~6300~\AA{} are common to be observed in relation with internally driven winds \citep[e.g.,][]{natta2014,simon2016,fang2018,nisini2018,banzatti2019,gangi2023} and also externally driven winds \citep[e.g.,][]{bally1998,Tsamis2013}. In distant clusters, the proplyd morphology cannot be spatially resolved. Furthermore, there may be no ionization fronts in regions where the EUV field is attenuated or negligible \citep[e.g.,][]{haworth2023}. These facts raise the question of how to identify external winds and how to disentangle them from internal ones. \citet{ballabio2023} explored the line luminosity of \OI\,6300~\AA{} as a diagnostic for external photoevaporation and predicted it to undergo a dramatic increase above $\sim$5000 G$_0$. However, it remains challenging to spectrally differentiate between internal and external winds without spatially resolving the system.

Here, we explore a proxy of externally driven winds, the near-infrared forbidden line of neutral atomic carbon at 8727~\AA{} covered by the MUSE-NFM data. Atomic carbon is predicted to be present in the upper layers of protoplanetary disks, where CO is dissociated by UV photons \citep[e.g.,][]{bruderer2012}. [C \textsc{i}]~609~$\mu m$ was recently detected in the isolated disk IM Lup with the Atacama Large Millimeter/submillimeter Array (ALMA) Band 8 observations, where it traces the disk atmosphere \citep{law2023}, and in the irradiated disk d203-506 in the ONC, where it traces a photoevaporative wind \citep{Goicoechea2024}. Forbidden carbon lines at 8727 \AA{} and 9850~\AA{} were observed in emission in the ONC before the discovery of proplyds (\citealt{hippelein1978} detected the two lines; \citealt{cosmovici1981} observed 8727~\AA). \citet{Goicoechea2024} also reported the detection of the \CI~9850~\AA{} line in d203-506. Other carbon emission lines around $\sim 1~\mu$m were observed in the Orion Bar by \citet{Walmsley2000} and \citet{peeters2024A&A...685A..74P}. Various permitted carbon emission lines in the same wavelength range were reported to be emitted in the inner regions of isolated disks by \citet{McClure2019} and \cite{McClure2020}, who did not detect near-infrared carbon forbidden lines in these objects. More recently, \CI~8727~\AA{} was detected in two irradiated disks in the ONC \citep{haworth2023}. This line, emitted at the 2.7~eV level in the transition 2$p^2$ $^1$D$_2$, was theoretically studied by \citet{escalante1991} together with two related forbidden lines at 9850 \AA{} and 9823 \AA. \citet{escalante1991} reported that these lines originate from recombination of C$^+$ ions produced by photoionization in regions with density larger than $10^5~{\rm cm}^{-3}$ and radiation fields $\sim 10^3-10^6 G_0$, where $G_0$ is the interstellar field intensity. More recent models predict that the line is instead excited via FUV-pumping and its intensity scales with $G_0$ \citep{Goicoechea2024}. 

In this Letter, we build on these previous works to present strong evidence that the [C \textsc{i}]~8727 \AA{} line can be used as a near-IR tracer for identifying externally photoevaporated disks combining the information on the location of the emission in this line and from the disk from 12 proplyds in the ONC observed with MUSE and with ALMA. 

\section{Data and analysis}\label{sect:data-and-analysis}

\subsection{Observational data}

\begin{table}
\begin{center}
\caption{Coordinates and projected separations of the sample of proplyds.}
\begin{tabular}{l|ccc}
\hline \hline
Proplyd & RA & DEC & $d$(UV source) \\
 & hh:mm::ss.s & dd:mm:ss.s & [pc] \\
\hline
154-346 & 05:35:15.44 & $-$05:23:45.55 & 0.068 \\
167-325 & 05:35:16.72 & $-$05:23:25.5 & 0.009  \\
168-326 & 05:35:16.85 & $-$05:23:26.22 & 0.012  \\
170-249 & 05:35:16.96 & $-$05:22:48.51 & 0.068 \\
170-334 &  05:35:16.96 & $-$05:23:33.6 & 0.028 \\
170-337 & 05:35:16.97 & $-$05:23:37.15 & 0.031 \\
171-340 & 05:35:17.06 & $-$05:23:39.77 & 0.037 \\
173-236 & 05:35:17.34 & $-$05:22:35.81 & 0.095 \\
174-414 & 05:35:17.40 & $-$05:24:14.5 & 0.106 \\
177-341W & 05:35:17.66 & $-$05:23:41.00 & 0.049 \\
203-504 & 05:35:20.26 & $-$05:25:04.05 & 0.077 ($\theta^2$ Ori A)\\
244-440 & 05:35:24.38 & $-$05:24:39.74 & 0.06 ($\theta^2$ Ori A)  \\
& & &  0.31  ($\theta^1$ Ori C)\\
\hline
\end{tabular}
\tablefoot{The information is from \citet{OdellWen1994,Bally2000DisksNebula,Ricci2008THENEBULA,Mann2014ALMAPROPLYDS}. For the given projected separations $d$, the UV source is $\theta^1$ Ori C with the exception of 203-504 (irradiated by $\theta^2$ Ori A) and 244-440.}
\label{table:coordinates}
\end{center}
\end{table}

We use data of the 12 proplyds showing prominent ionization fronts firstly presented by \citet{aru2024}. These objects were observed with the VLT/MUSE integral-field spectrograph \citep{Bacon2010}. The coordinates of the targets are given in Table~\ref{table:coordinates}. MUSE was operated in NFM, which covers the wavelength range $\sim$4750--9350 \AA\ with a field of view (FOV) of $\sim$7.5" $\times$ 7.5". The observations were taken in three programs (Pr. ID 104.C-0963(A) and 106.218X.001, PI: C. F. Manara; Pr. ID 110.259E.001, PI: T. J. Haworth). We measure the angular resolution of the images in the range FWHM$\approx$0.06-0.08" at $\sim$8760~\AA{} in our observations. 
Details regarding the observation and the data reduction process are described in \citet{haworth2023} for proplyd 203-504, and in \citet{aru2024} for the remainder of the sample. \\
In addition, we use the ALMA observations of six targets (168-326, 170-249, 170-337, 171-340, 173-236, 177-341W) performed in Band 7 (0.86 mm) with an angular resolution of 0.09". The ALMA Band 7 data trace thermal dust emission from the disk, and they are described in \citet{eisner2018}. \citet{ballering2023} show that the Band 7 data does not trace the ionization front, which is instead traced by the Band 3 data as presented in their work.  
Lastly, we use the publicly available spectra taken with VLT/X-Shooter for the PENELLOPE Large Program \citep{manara2021} and those published by \citet{mauco2023}.

\subsection{Data analysis}\label{sect:analysis}
In the following analysis, we consider the \OI\, lines at 5577~\AA{} and 6300~\AA\ and \CI\,8727~\AA, observed with MUSE-NFM, and compare the location of the continuum-subtracted emission (see \citealt{aru2024} for details) with the locations of the ALMA disk continuum emission and the stellar continuum emission. 

\subsubsection{Alignment and spatial comparison}\label{analysis:alignment}

As MUSE-NFM is known to have uncertain astrometric accuracy, it is necessary to align the MUSE cubes with the ALMA data before analyzing how the MUSE detections of \CI\,8727 \AA{}, and \OI\,5577~\AA{} and 6300~\AA{} emission compare to the disk dust observations of ALMA. The alignment was done by measuring relevant references points in the MUSE and ALMA continuum images and shifting the coordinates of the ALMA image to align with MUSE. 

In the case of 177-341W, we carried out the matching by using two background sources present in both the ALMA and MUSE data. We refer to Appendix~\ref{apx-sect:177-341} for a more detailed discussion on this target and its alignment. 
For the remaining proplyds, we used a subcube of the MUSE observations nearly free of emission lines, showing the star continuum, at $\lambda\sim674-678$~nm. We then aligned the location of the star, measured on the MUSE image, with the center of the disk in the ALMA continuum image. We considered the uncertainty on this alignment to be half the spatial size of the MUSE spaxels (0.0125") for all proplyds except 168-326, 174-414 and 203-504 for which the spaxel size 0.025" was used as the emission around the central star is more extended and therefore the location more uncertain.

\begin{figure}
    \centering
    \includegraphics[width=0.99\columnwidth]{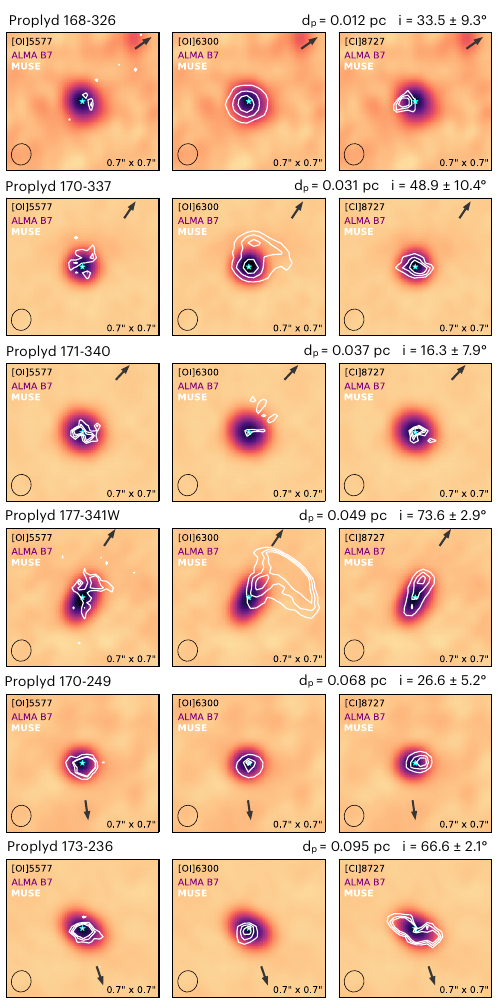}
    \caption{ALMA Band 7 images of proplyds 168-326, 170-337, 171-340, 177-341W, 170-249, and 173-236 compared with the MUSE emission lines (contours) at 50\%, 70\% and 90\% of the peak intensity; the rows are ordered in increasing projected distance from \thetaonec. The ALMA Band 7 data is originally published by \citet{eisner2018}; the values for inclination marked above the figures are from \citet{ballering2023}, except for 168-326 and 171-340 (Ballering et al., in prep.). The MUSE emission line and the size of each image are shown in the top left corner, and the beam size is indicated in the bottom left corner. The direction of the UV source is shown with an arrow. The cyan star marks the estimated star's location.}
    \label{fig:alma-muse}
\end{figure}

In Figure~\ref{fig:alma-muse}, the images of the ALMA Band 7 data are shown together with the line emission from MUSE, which is shown with contours representing 50\%, 70\% and 90\% of the peak intensity of the line emission.  
The proplyds are ordered starting from the smallest projected separation to \thetaonec\, to the largest. The direction of the UV source is shown with an arrow. We report the inclinations of the disks, calculated from their deconvolved FWHM major and minor axes listed by \citet{ballering2023} and calculated using the CASA task imfit on the ALMA data for proplyds 168-326 and 171-340.

\begin{figure*}
    \centering
    \includegraphics[width=\textwidth]{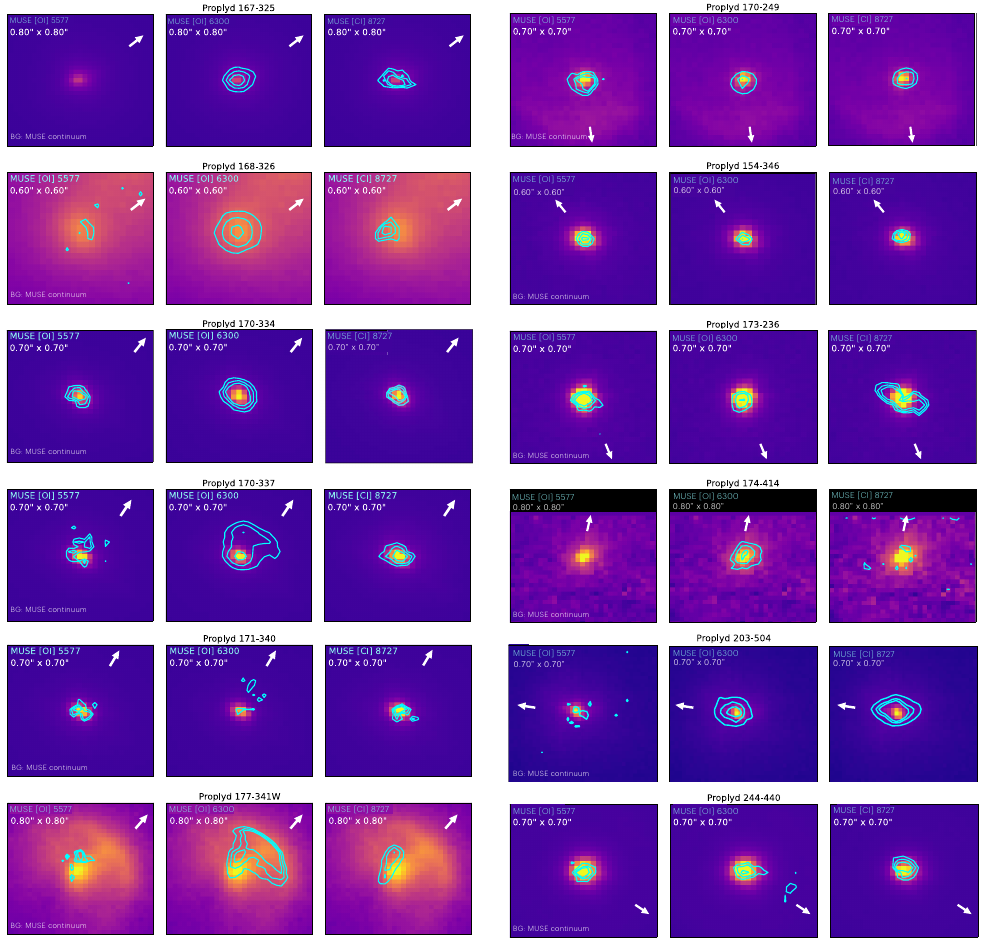}
    \caption{Spatial comparison between MUSE stellar continuum (colormap) and MUSE emission lines (contours) for proplyds 154-346, 167-325, 168-326, 170-249, 170-334, 170-337, 171-340, 173-236, 174-414, 177-341W, 203-504, and 244-440. The emission line and the size of each image are shown in the top left corner. The contours represent 50\%, 70\% and 90\% of the peak intensity of the MUSE emission lines, except for the \CI\, line of 174-414 for which only 90\% is shown. The direction of the UV source is shown with an arrow. }
    \label{fig:muse-muse}
\end{figure*}

We made a spatial comparison between the stellar emission location and the \OI\,5577~\AA, 6300~\AA{} and \CI\,8727~\AA{} line emission location to determine any spatial displacement.
Figure~\ref{fig:muse-muse} shows the MUSE stellar continuum emission and the contours of the emission lines similarly to Figure~\ref{fig:alma-muse}. In Figure~\ref{fig:muse-muse} all the 12 proplyds are shown, including those not observed with ALMA. 

\subsubsection{Radial cuts}

In order to confirm that the line emission is spatially resolved and analyze the shift of the emission lines with respect to the central star, we retrieved a radial cut for each proplyd in the MUSE emission line images. 
This was performed by calculating the emission along a 6" long line centered at the coordinates of the central star and oriented towards the ionizing source \thetaonec.
The procedure is described in more detail in \citet{aru2024}. In the case of 177-341W, we centered the radial cut line instead at coordinates corresponding to the center of the disk as seen in the ALMA image. This is because the disk is highly inclined, and the coordinates of the central star cannot be determined in the same way as for other proplyds. This target is discussed in more details in Appendix~\ref{apx-sect:177-341}. 
Figures of radial cuts are shown in Appendix \ref{apx-fig:radialcuts}.

\begin{figure}
    \centering
    \includegraphics[width=\columnwidth]{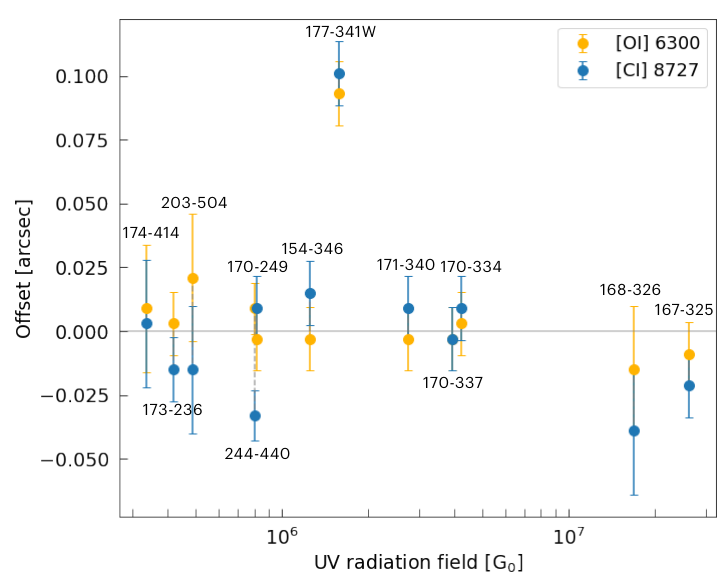}
    \caption{Offsets between the intensity peaks of forbidden line emission and the location of the star versus the UV radiation field. The peak in the radial cut of  \OI~6300~\AA\ is marked with orange, and blue points mark \CI~8727~\AA.}
    \label{fig:g0-offsets}
\end{figure}

\section{Results and discussion}\label{sect:discussion}
Here we describe the observed spatial location of the \OI\,5577~\AA, 6300~\AA{} and \CI\,8727~\AA{} lines observed with MUSE, and how these compare with the disk emission traced by ALMA. Combining this information with analysis of high-resolution spectra of young stars in low-mass star-forming regions, we show the potential of the \CI{} line to trace externally photoevaporated winds.

\subsection{[OI]~5577~\AA, 6300~\AA{} lines in proplyds}

The \OI\,6300 \AA{} line was detected in some of the proplyds in the ONC with HST \citep[e.g.,][]{bally1998}, while \OI\,5577 \AA{} was detected in proplyd 182-413 (also known as HST-10) with VLT/FLAMES \citep{Tsamis2013}. 
We detect both of the lines in all 12 proplyds of the sample. Here, we introduce a comparison between these oxygen emission lines and the ALMA disk continuum emission. 

The location of the \OI\,5577~\AA, 6300~\AA{} emission lines is typically well aligned with the dust disk emission from ALMA, as shown in Figure \ref{fig:alma-muse}. The emission of \OI\,5577~\AA{} coincides with the disk, which is seen the most clearly, due to the lowest noise, in proplyds 170-249 and 173-236. We note that these are the two disks with largest projected separation from the ionizing source in the combined MUSE-ALMA sample shown here. In the MUSE-MUSE comparison of Figure~\ref{fig:muse-muse}, proplyds 170-334, 154-346, 203-504, and 244-440 also show the coincidence of the \OI\,5577~\AA{} emission with the disk, under the assumption that the star is centered at the center of the disk. For 167-32 and 174-414, the MUSE moment 0 map is too noisy for \OI\,5577~\AA{} contours to be shown.\\
In a study of the proplyd 182-413 \citep{Tsamis2013}, the line \OI\,5577 \AA{} is found to peak at spaxels which coincide with the position of the disk, therefore located similarly to our observations. We thus confirm that the \OI\,5577~\AA{} emission is originated on the disk surface.

The \OI\,5577~\AA{} emission is not co-spatial with \OI\,6300~\AA. The major difference between the two oxygen emission lines is that \OI\,5577~\AA{} is only bound to the disk region, and it is not emitted in the ionization front. The \OI~6300~\AA{} emission from the disk and the ionization front is connected in the contours for proplyds 170-337 and 177-341W; these locations match with the early observations described by \citet{bally1998}. Theoretical works have also predicted \OI\,6300~\AA{} to trace multiple regions: the area near the disk/the disk surface, the photodissociation region, and the ionization front where oxygen is excited thermally due to collisions with H, H$_2$ and electrons \citep[e.g.,][]{storzer-hollenach-1998}. 

In Figure~\ref{fig:g0-offsets}, we show the displacement of the \OI\,6300 \AA{} peak emission from the location of the star as a function of the UV radiation field. The positive sign of the values on y-axis marks a peak located in the direction of the UV source, and the negative sign the opposite. These values are derived from the radial cuts shown in Figure~\ref{apx-fig:radialcuts}, and given in Table~\ref{table:measurements}.  
The values are compatible with no displacement within their uncertainties, except for the case of 177-341W, an inclined disk of $i$=73.6~$\pm$~2.9~deg \citep{ballering2023}. 

\begin{table*}
\begin{center}
\caption{Coordinates and projected separations of the detected proplyds.}
\begin{tabular}{l|ccccc}
\hline \hline
Proplyd &  Offset, \OI\, (") &  Offset, \CI\, (") & i (deg) &  PA$_d$ (deg) & PA$_{d,\star}$ (deg) \\
(1) & (2) & (3) & (4) & (5) & (6) \\
\hline
154-346 & -0.003 $\pm$ 0.0125 & 0.015 $\pm$ 0.0125 & \nodata & \nodata & 38 \\
167-325 & -0.009 $\pm$ 0.0125 & -0.021 $\pm$ 0.0125 & \nodata & \nodata & 126 \\
168-326 &    -0.0150 $\pm$ 0.0250 &     -0.039 $\pm$ 0.0250 &       33.5 $\pm$ 9.3 &  31 $\pm$ 3 & 125 \\
170-249 &    -0.0030 $\pm$ 0.0125 &      0.009 $\pm$ 0.0125 &       26.6 $\pm$ 5.2 &       68 $\pm$ 16  & 10 \\
170-334 &   0.003 $\pm$ 0.0125 & 0.009 $\pm$ 0.0125 & \nodata & \nodata & 136 \\ 
171-340 &    -0.0030 $\pm$ 0.0125 & 0.009 $\pm$ 0.0125 & 16.3 $\pm$ 7.9 &       41 $\pm$ 9 & 148 \\
170-337 &    -0.0030 $\pm$ 0.0125 &     -0.003 $\pm$ 0.0125 &       48.9 $\pm$10.4 &       85 $\pm$ 17 & 147 \\
173-236 &     0.0030 $\pm$ 0.0125 &     -0.015 $\pm$ 0.0125 &       66.6 $\pm$ 2.1 &       60 $\pm$  2 & 18 \\
174-414 & 0.009 $\pm$ 0.0250 & 0.003 $\pm$ 0.0250 & \nodata & \nodata & 164 \\ 
177-341W &     0.0931 $\pm$ 0.0125 &      0.101 $\pm$ 0.0125 &       73.6 $\pm$ 2.9 &      152 $\pm$ 2 & 136 \\
203-504 & 0.021  $\pm$  0.0250 & -0.015 $\pm$  0.0250 & \nodata & \nodata & 75 \\ 
244-440 & 0.009 $\pm$ 0.01 & -0.033 $\pm$ 0.01 & \nodata & \nodata & 52  \\ 
\hline
\end{tabular}
\tablefoot{Column (1): name of the proplyd. Column (2): the peak of \OI\,~6300~\AA{} emission measured from the location of the star in a radial cut; the sign indicates whether the peak is towards or in the opposite side from the UV source. Column (3): similar as column (2) but for \CI\,8727~\AA{}. (4): inclination of the disk, calculated from the deconvolved FWHM major and minor axes of disks listed by \citet{ballering2023}, and Ballering et al., in prep. for proplyds 168-326 and 171-340. Column (5): position angle of the deconvolved disk from ALMA data similarly to column (4). Column (6): position angle between the disk and the UV source. As the inclination and PA$_d$ are measured from ALMA data, the values are not available for the whole sample. }
\label{table:measurements}
\end{center}
\end{table*}

To investigate further whether the offset of the emission peak is an effect related to the relative disk inclination with respect to the UV source, we calculate the relative inclination of the disks using the spherical law of cosines: $\cos(\Delta i)=\cos(i_1)\cos(i_2) + \sin(i_1)\sin(i_2) \cos(\Omega_1 - \Omega_2)$, where $\Omega_1$ is the position angle of the disk from ALMA, $\Omega_2$ the position angle between the central star and the UV source, $i_1$ the disk inclination, and $i_2=90$~deg is the inclination we assume for the emission coming from the UV source in the plane of the sky.
In Figure~\ref{fig:rel-inc-offsets}, the displacements are shown as a function of the relative inclination. The proplyd 177-341W has the highest relative inclination, which could indeed explain that the displacement arises from the disk's inclination with respect to \thetaonec. Indeed, if the \OI{} emission is mainly coming from a wind on the surface of the disk, it is to be expected that this is mainly arising from the edge of the disk on the side pointing towards the UV source \thetaonec. In cases of low relative inclination, this effect is more difficult to see as the whole disk surface is irradiated, resulting in a more homogeneous emission. On the contrary, when the relative inclination is higher, then one side of the disk is more illuminated and the emission is stronger. This hypothesis needs confirmation on a larger sample of targets.

\subsection{[CI]~8727~\AA{} line in proplyds}

The MUSE data revealed the first detection of the \CI\,8727~\AA{} emission line in proplyds \citep{haworth2023}. Here, we present for the first time a comparison between \CI\,8727~\AA{} emission and the ALMA dust emission, and the detection of this line in all 12 proplyds in our sample. 

Firstly, the \CI\,8727~\AA{} emission appears co-spatial with the ALMA disk emission. No contribution from the ionization front is present, and the line is emitted only from the disk surface and/or base of the externally photoevaporating wind. 
Compared to the contours of \OI\,$\lambda$5577, $\lambda$6300 lines, \CI\,8727~\AA{} traces the disk shape more closely in the case of inclined disks 173-236 and 177-341W, as seen in Figure \ref{fig:alma-muse}. In the case of 173-236, the contours of \CI\,8727~\AA{} follow an elliptical shape which coincides with the ALMA continuum image, compared to the circular contours of the \OI\,6300
\AA{} emission line. The shape of the \CI\,8727~\AA{} contours is also elliptical for 177-341W, whereas \OI\,5577~\AA{} is not clear and the contours of \OI\,6300~\AA{} are merged with the emission from the ionization front. For 170-337, \CI\,8727~\AA{}, also pinpoints to the disk without noise or emission from the ionization front. Therefore, \CI\,8727~\AA{} is a more accurate tracer of the disk surface and/or the base of the externally photoevaporating wind than the \OI\, lines. Furthermore, the similarity between the gas radius and the dust continuum radius could potentially allow the \CI\,8727~\AA{} emission line to be used for estimating the gas disk radius. 

\begin{figure}
    \centering
    \includegraphics[width=\columnwidth]{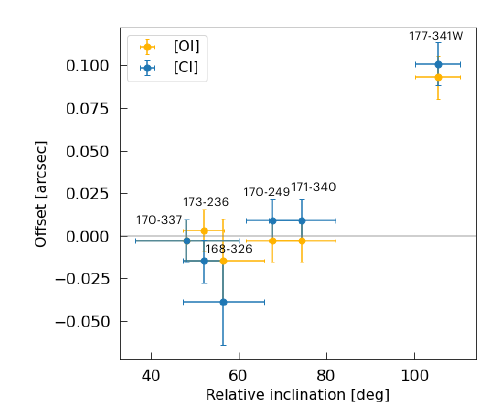}
    \caption{Offsets between the intensity peaks of forbidden line emission and the location of the star versus relative inclination. A relative inclination of 0 deg would correspond to a disk being irradiated face-on, and of about 90 deg to an edge-on configuration. The peak in the radial cut of  \OI~6300~\AA\ is marked with yellow, and blue points mark the offset of \CI~8727~\AA. }
    \label{fig:rel-inc-offsets}
\end{figure}

\begin{figure*}
    \centering
    \includegraphics[width=\textwidth]{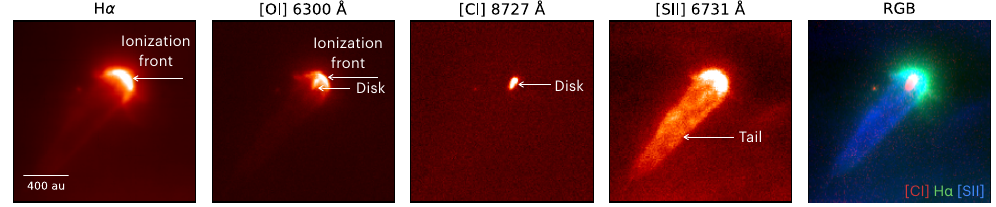}
    \caption{Continuum subtracted, single line integrated flux images and an RGB image of proplyd 177-341W. The panels show which parts of the system are visible in given emission lines.}
    \label{fig:177-341-collage}
\end{figure*}

The \CI\,8727~\AA{} emission is co-spatial with the MUSE stellar continuum emission for the rest of the proplyds in the sample (170-334, 174-414, 167-325, 154-346, 203-504, and 244-440). The stellar continuum, taken in an emission-line free region of the MUSE cube between 8756--8764~\AA, is shown in Figure~\ref{fig:muse-muse} together with the contours showing the location of the emission lines. 
When comparing the peak emission offset with the irradiation from the ionizing source on the targets (Figure~\ref{fig:g0-offsets}), also for the \CI\,8727~\AA{} line the biggest displacement is seen in 177-341W. 
For 177-341W, the \CI\,8727~\AA{} emission peaks on the side of the disk facing towards the direction of the UV source. The offset between the center of the disk in the ALMA image and the peak of \CI\,8727~\AA{} emission (MUSE) is shown as an outlier in Figure~\ref{fig:rel-inc-offsets}, as the location of the star was assumed to be in the center of the disk. This peak implies that \CI\,8727~\AA{} traces the surface of the disk and/or the base of the photoevaporative wind. In a similar way as for the discussion on the \OI\, line offset, the largest offset in this case may be due to the high relative inclination of this disk with respect to \thetaonec. The targets with lower relative inclinations show a more uniform emission on the disk surface. The lack of measurements of the disk inclination and position angle for half of our sample hinders the possibility to further confirm this is due to the relative inclination of the disks and the external UV radiation. An example of the proplyd morphology based on 177-341W as observed with MUSE is shown in Figure~\ref{fig:177-341-collage}.

Proplyd 168-326 is another outlier in Figure~\ref{fig:g0-offsets}-\ref{fig:rel-inc-offsets} with the peak of \CI\,8727~\AA{} facing away from \thetaonec. The proplyd has the second smallest projected separation ($d_p$=0.012~pc) from the UV source among the 12 targets. However, the contours of its nearby proplyd 167-325 ($d_p$=0.009~pc) coincide with the location of the central star, as seen in Figure~\ref{fig:muse-muse}. Therefore, it is unclear whether the displacement of \CI\,8727~\AA{} could be explained due to the small projected separation. It is also to be noted that, for this target, the offset is still compatible with zero, and therefore it is harder to conclude any statistically sound difference. 

Finally, we note that also 244-440 has a peak offset of the \CI\,8727~\AA{} line slightly larger than zero and in the opposite direction of the ionizing radiation (Figure~\ref{fig:g0-offsets}). We note however that this is a peculiar target possibly irradiated by two UV sources \citep{ODell2017WhichNebula}. The properties of this proplyd are discussed in more detail in \citet{aru2024} and it will be subject of future studies. 

\subsection{The [CI] 8727~\AA{} line as a tracer of externally photoevaporated disks}

The spatial location of the \CI\,8727~\AA{} emission strongly suggest that this line is originated on the disk surface and/or the base of the externally photoevaporating wind. It is interesting to verify whether this line is tracing disk winds also in low-ionization environments, where external photoevaporation is not at act. This is the case for the \OI\, lines, routinely detected in isolated disks in low-mass SFRs, and interpreted in those environments as tracers of internal - photoevaporative or magneto-hydro dynamic (MHD) - winds \citep[][for a review]{pascucci2023}. 

To verify whether the \CI\,8727~\AA{} line is also typical in disks in low-mass SFRs where external photoevaporation is not expected, we examined 74 VLT/X-Shooter spectra from the PENELLOPE Large Program \citep[][regions such $\epsilon$ Cha, Lupus, Taurus]{manara2021} and 45 VLT/X-Shooter spectra from the study of $\sigma$-Orionis by \citet{mauco2023}. The \OI\,5577~\AA{} line is detected in 68\% of the targets, and the \OI\,6300~\AA{} line in 88\% of the targets in the PENELLOPE sample (Campbell-White, in prep.). In contrast, only one target shows a clear detection for the \CI\,8727~\AA{} line in the PENELLOPE sample, with 7 additional tentative detections, for a total of $\ll10\%$ detection rate. This number increases substantially to a $\sim40\%$ detection rate in $\sigma$-Orionis, a region with a higher UV radiation than in low-mass SFRs, but lower than in ONC. It is to be expected that a good fraction of the targets in this region are indeed experiencing external photoevaporation winds. Finally, we have detected this line in all 12 proplyds in the ONC observed with MUSE/NFM. The lack of detections in isolated disks is consistent with \citet{McClure2019} and \citet{McClure2020}, who did not detect infrared carbon forbidden lines in the innermost regions of isolated disks, where instead they detected several permitted carbon lines. In the same disks, previous works had detected forbidden lines of \OI{}\,in the vast majority of the targets \citep{natta2014,nisini2018}. 

This notable difference in detection rates between the \OI\, optical lines and the \CI\,8727~\AA{} lines in disks in low-mass star-forming regions strongly points to the fact that this line is emitted in highly externally irradiated environments. This is in agreement with the results of \citet{escalante1991}, who point to high external UV flux in cluster environments, like the one experienced by proplyds, for this line to be emitted.
We note that they considered the $10^3$--$10^6$~$G_0$ FUV field range, and the FUV field of our targets expands this range into lower values, with $10^2$~$G_0$ in $\sigma$-Orionis, and higher range for the proplyds closest to \thetaonec\,in the ONC (167-325 and 168-326, $10^7$~$G_0$).
Our results are even in better agreement with the FUV-pumping excitation models of \citet{Goicoechea2024}. Their models also predict that the intensity of the IR carbon lines scales with $G_0$. This dependence should be explored in future works. As expected from these models, the emission of \CI\,8727~\AA{} is confined to the disk and/or base of the externally photoevaporating wind, and does not appear on the ionization front, in agreement with the fact that these are not recombination lines. 
The detection rate, together with the coincidence of the emission with the disk surface strongly suggest that the \CI\,8727~\AA{} line is emitted in an externally photoevaporated wind, and does not suffer from confusion with line emission due to internal processes, like in the case of the \OI\, lines, known tracers of internal processes \citep[e.g.,][]{ercolano2017} and external winds \citep[e.g,][]{ballabio2023}. 

\section{Conclusions}\label{sect:conclusion}

In this work, we compared the spatial location traced by VLT/MUSE-NFM of the emission in the \OI\,6300~\AA{} and \OI\,5577~\AA{} lines, as well as in the \CI\,8727~\AA{} line, with the ALMA Band 7 continuum disk emission in a sample of 12 proplyds in the ONC.

We confirmed that the \OI\, emission is co-spatial with the disk emission in the case of the \OI\,5577~\AA{} line, whereas the \OI\,6300~\AA{} is emitted both on the disk surface and on the ionization front of the proplyds. At the same time, we showed for the first time that \CI\,8727~\AA{} is also co-spatial with the disk surface in a proplyd, as seen in the MUSE and ALMA data comparison. The peak emission is compatible with the stellar location in all cases, apart from one target with high relative inclination with respect to the UV radiation, where the peak emission is located at the disk edge in the direction of the UV radiation. 

 We have presented the [C \textsc{i}] 8727~\AA{} emission line in 12 proplyds that were observed with MUSE in NFM, with a detection rate of 100\%. In contrast, this line has a much lower detection rate in other SFRs, which goes down to be compatible with zero detections in the nearby low-mass SFRs. This result strongly support the recent excitation models of neutral carbon in externally irradiated disks proposed by \citet{Goicoechea2024}. 
On the other hand, [O \textsc{i}] lines at 5577 \AA{} and 6300~\AA{} are common in low mass SFRs, as they are also related to internal disk winds, and this contamination makes them a less direct tracer of externally driven winds.

Our results strongly suggest that:

\begin{itemize}
    \item The \CI\,8727~\AA{} line is tracing the base of the externally photoevaporated wind, can act as a key tracer of the wind, and is ideal to distinguish from winds in disks which are not externally photoevaporated, such as in low-mass SFRs.
    \item This emission can be particularly valuable for identifying external irradiation when there is no ionization front/proplyd morphology visible (as in the case of d203-506, \citealt{haworth2023}).
    \item As the dust continuum radius is very similar to the gas radius, the \CI{} line could be a potential tracer of the gas disk radius (and by extension, the gas-disk size distribution) in irradiated environments, to be explored in future works. 
\end{itemize}

Our work highlights the strength of using the \CI\,8727~\AA{} line as a proxy to trace external photoevaporation.
With VLT/MUSE, it may be systematically easier to detect large samples of externally photoevaporated gas disks in massive clusters, and to study the global spatial extents of the dust vs. gas at large samples in combination with ALMA. 

Future studies should aim at studying the kinematics of this line by using spectral resolutions inaccessible with MUSE-NFM, to confirm that this line is tracing a slow wind, as expected from externally photoevaporated winds \citep{ballabio2023}. Additional surveys with ALMA should be carried out to measure the disk inclination and position angles, as well as the disk sizes, in a larger number of proplyds, to be then combined with the MUSE data to further decipher their physical conditions. 

\begin{acknowledgements}
We thank the anonymous referee for detailed comments and useful suggestions, which improved our work.
MLA, KM, CFM, and JCW are funded by the European Union (ERC, WANDA, 101039452). GPR is funded by the European Union (ERC, DiscEvol, 101039651) and by the Fondazione Cariplo, grant no. 2022-1217. 
SF is funded by the European Union (ERC, UNVEIL, 101076613), and acknowledges financial contribution from PRIN-MUR 2022YP5ACE. TJH acknowledges funding from a Royal Society Dorothy Hodgkin Fellowship and UKRI guaranteed funding for a Horizon Europe ERC consolidator grant (EP/Y024710/1). GB is supported by the European Research Council (ERC) under the European Union’s Horizon 2020 research and innovation programme (Grant agreement No. 853022, PEVAP). 
 
Views and opinions expressed are however those of the author(s) only and do not necessarily reflect those of the European Union or the European Research Council Executive Agency. Neither the European Union nor the granting authority can be held responsible for them.\\
This paper makes use of the following ALMA data: ADS/JAO.ALMA \#2015.1.00534.S. ALMA is a partnership of ESO (representing its member states), NSF (USA) and NINS (Japan), together with NRC (Canada), NSTC and ASIAA (Taiwan), and KASI (Republic of Korea), in cooperation with the Republic of Chile. The Joint ALMA Observatory is operated by ESO, AUI/NRAO and NAOJ.

\end{acknowledgements}

\bibliographystyle{aa} 
\bibliography{ref} 

\appendix

\section{The case of proplyd 177-341W}\label{apx-sect:177-341}

\begin{figure*}
    \centering
    \includegraphics[width=0.99\textwidth]{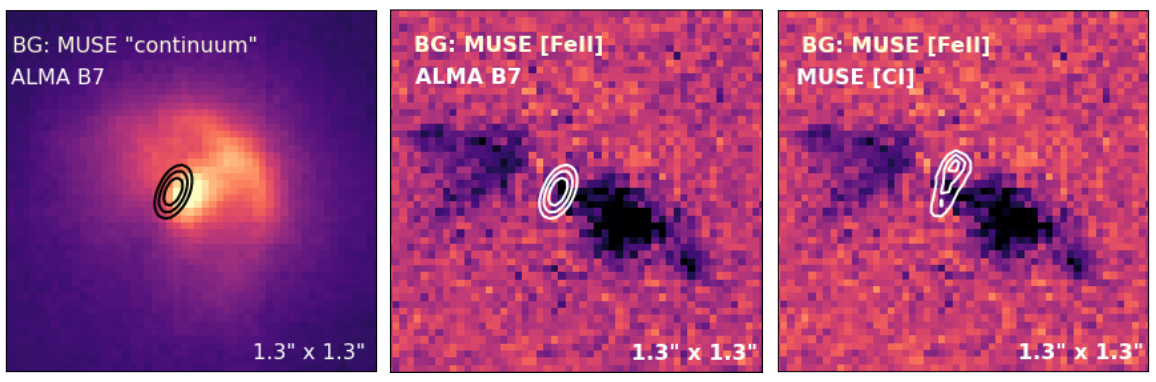}
    \caption{Proplyd 177-341W shown for three cases: 1) the stellar continuum emission in the background, overlaid with the contours of ALMA Band 7 data; 2) [Fe \textsc{ii}]~8892~\AA{} (MUSE) in the background, overlaid with the contours of ALMA Band 7 data; 3) [Fe \textsc{ii}]~8892~\AA{} (MUSE) in the background, overlaid with the contours of \CI\,8727 (MUSE). }
    \label{apx-fig:177-341}
\end{figure*}

With Figure~\ref{apx-fig:177-341}, we investigate the location of the disk (observed in ALMA band 7) with respect to the subcube nearly free of emission lines (left-most panel) and the jet as seen in the emission line [Fe \textsc{ii}]~8892~\AA{} observed with MUSE (middle panel). We also show the \CI\, emission observed with MUSE as overlaid on the moment 0 map showing the jet (right-most panel).

The jet pinpoints to the center of the disk and therefore the location of the star, further indicating that the peak of the emission in the left-most panel is misaligned from the true location of the star. The case of 177-341W is similar to the scattered light images of highly inclined disks studied by \citet{villenave2020}.

\section{Additional figures}\label{apx-sect:additional}

\begin{figure*}
    \centering
    \includegraphics[width=0.95\textwidth]{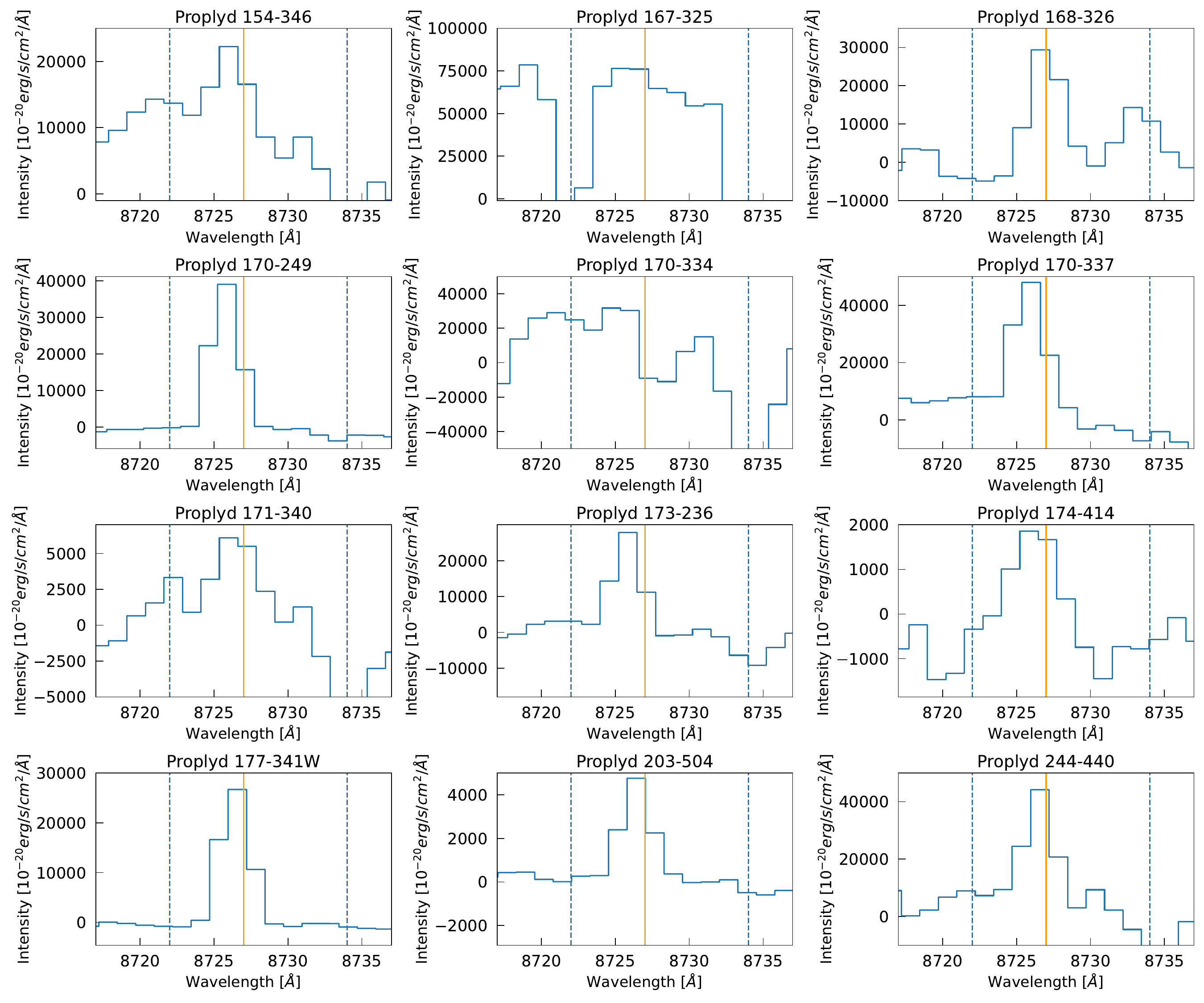}
    \caption{The spectra of 12 proplyds in the wavelength range 8717--8737 \AA. The orange line marks 8727 \AA, and the dashed lines mark the range in which the MUSE \CI\, emission line image was created.}
    \label{apx-fig:ci-spectra}
\end{figure*}

In Figure~\ref{apx-fig:ci-spectra}, the \CI\, line at 8727~\AA{} in the 12 proplyds is shown. The spectra were extracted from continuum subtracted cubes of wavelength range 8660--8760~\AA{} on the emission, by using a circular aperture of 0.1". 

\begin{figure*}
    \centering
    \includegraphics[width=0.95\textwidth]{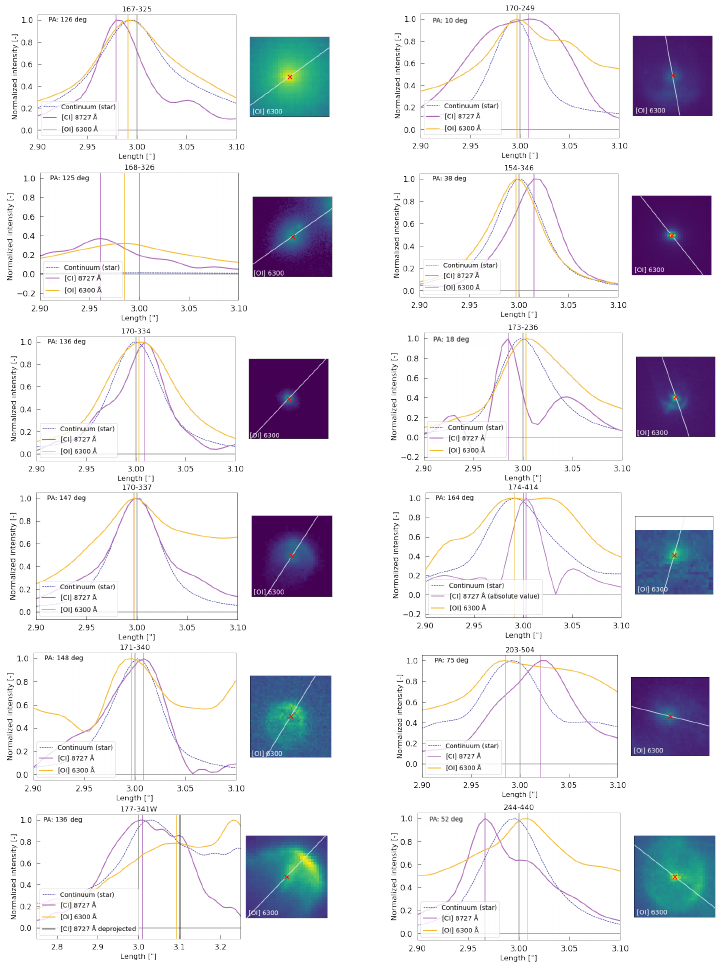}
    \caption{Radial cuts of MUSE-NFM images taken at the position angle between the central star and the UV source; the method is illustrated with image panels, where the coordinates of the central star are marked with X. Three lines show the radial cuts of \OI~6300~\AA, \CI~8727~\AA, and stellar continuum images. }
    \label{apx-fig:radialcuts}
\end{figure*}

Figure~\ref{apx-fig:radialcuts} shows the radial cuts of 12 proplyds for the MUSE stellar continuum, and the moment 0 maps of \OI\,5577~\AA{} and \CI\,8727~\AA{} emission. These radial profiles were taken in the direction of the UV source, and allowed us to measure the peak of the emission lines with respect to the location of the star.

\begin{figure*}
    \centering
    \includegraphics[width=0.8\textwidth]{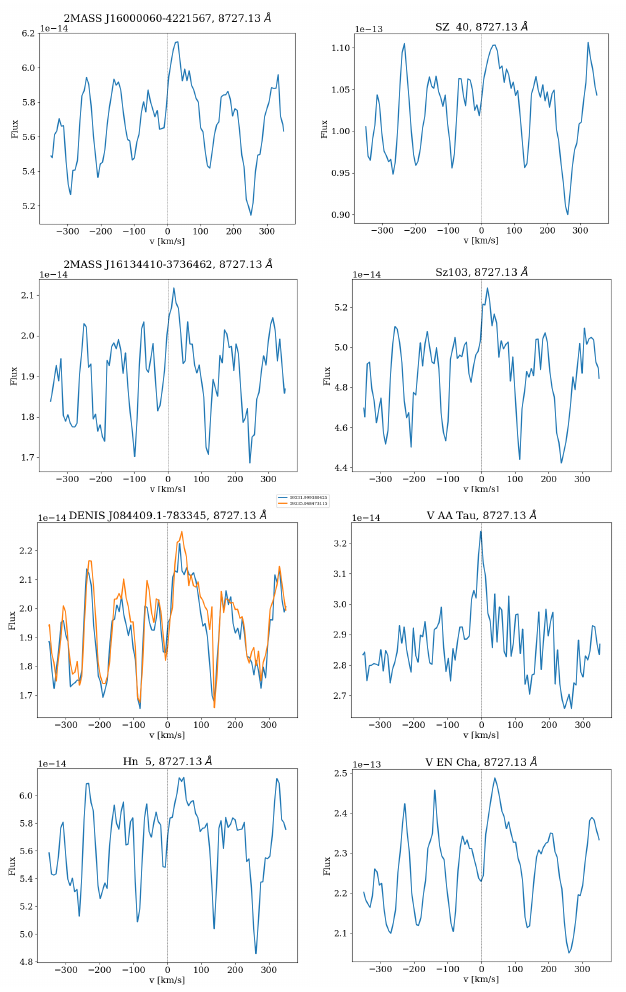}
    \caption{Spectra of the eight tentative detections of the \CI\,8727 line out of 74 targets observed with VLT/X-Shooter. The spectra are corrected for barycentric velocities.}
    \label{apx-fig:lowmass-spectra}
\end{figure*}

Figure~\ref{apx-fig:lowmass-spectra} shows the spectra of eight tentative detections of the \CI\,8727 line, out of 74 targets observed with VLT/X-Shooter. The clearest detection is seen in the case of AA Tau.

%
%

\end{document}